\documentclass[aip,jap,reprint,graphicx,twocolumn]{revtex4-1}%
\usepackage{epsf}
\usepackage{graphicx}
\usepackage{color}
\usepackage{amsmath}
\usepackage{amsfonts}
\usepackage{amssymb}%
\providecommand{\U}[1]{\protect\rule{.1in}{.1in}}
\begin{document}

\title{Current-induced spin torque resonance of magnetic insulators affected by
field-like spin-orbit torques and out-of-plane magnetizations}
\date{\today}
\author{Takahiro Chiba}
\email{t.chiba@imr.tohoku.ac.jp}
\affiliation{Institute for Materials Research, Tohoku University, Sendai, Miyagi 980-8577, Japan}
\author{Michael Schreier}
\affiliation{Walther-Meissner-Institut, Bayerische Akademie der Wissenschaften, Walther-Meissner-Strasse 8, 85748 Garching, Germany}
\author{Gerrit E. W. Bauer}
\affiliation{Institute for Materials Research, Tohoku University, Sendai, Miyagi 980-8577, Japan}
\affiliation{WPI-AIMR, Tohoku University, Sendai, Miyagi 980-8577, Japan}
\affiliation{Kavli Institute of NanoScience, Delft University of Technology, Lorentzweg 1,
2628 CJ Delft, The Netherlands}
\author{Saburo Takahashi}
\affiliation{Institute for Materials Research, Tohoku University, Sendai, Miyagi 980-8577, Japan}

\begin{abstract}
The spin-torque ferromagnetic resonance (ST-FMR) in a bilayer system
consisting of a magnetic insulator such as $\mathrm{Y_{3}Fe_{5}O_{12}}$ and a
normal metal with spin-orbit interaction such as Pt is addressed theoretically.
We model the ST-FMR for all magnetization directions and in the presence of
field-like spin-orbit torques based on the drift-diffusion spin model and
quantum mechanical boundary conditions. ST-FMR experiments may expose crucial
information about the spin-orbit coupling between currents and magnetization
in the bilayers.
\end{abstract}
\maketitle

\section{Introduction}

Near-dissipationless propagation of spin waves utilizing magnetic insulators
such as $\mathrm{Y_{3}Fe_{5}O_{12}}$ (YIG) with very low magnetization damping
by Pt contacts creates an interface between spinelectronic and magnonic
circuits for low power data transmission.\cite{Kajiwara,Wu} However, the
interpretation of experiments of current-induced coherent spin waves or
magnetization dynamics is unclear by the strongly non-linear proses with a
problematic threshold.\cite{Xiao} Spin wave mediated transport in YIG$|$Pt
bilayer has only recently been discovered and attracted a great deal of
interest since it revealed new physics by, \textit{e.g.}, the spin Seebeck
effect (SSE)\cite{Uchida10} and spin Hall magnetoresistance (SMR),
\cite{Nakayama,Chen} implying application potential for low-dissipation
spintronic interconnects and large area thermoelectric power generation. The
SMR refers to the dependence of the electrical resistance of the normal metal
on the magnetization angle of an adjacent magnetic insulator and is caused by
a simultaneous operation of the Spin Hall Effect (SHE) \cite{review12}
and its inverse (ISHE) as a nonequilibrium proximity phenomenon. Many
experiments have been described quantitatively well by the SMR model with one
set of parameters.\cite{Althammer,Hahn,Vlietstra,Marmion,Lin,Isasa} The magnetoresistance enables
straightforward access to the effects of spin-orbit coupling between currents
and magnetization in the bilayers.

The spin current through a ferromagnet$|$normal metal interface is governed by
the complex spin-mixing conductance (per unit area of the interface)
$G^{\uparrow\downarrow}=G_{r}+iG_{i}$. \cite{Tserkovnyak05} The prediction of
a large $G_{r}$ for interfaces between YIG and simple metals by
first-principle calculations \cite{Jia11} has been amply confirmed by recent
experiments.\cite{Weiler} The imaginary part $G_{i}$ can be interpreted as an
effective exchange field between magnetization and spin accumulation, which in
the absence of spin-orbit interaction is usually much smaller than the real
part. However, in metallic structures field-like spin-orbit torques (SOTs)
have been found, which can be modelled by a significant $G_{i}$%
.\cite{Pai14,Kim14} Current-induced SOTs are often associated with the Rashba
spin-orbit interaction.\cite{Haney,Skinner} In the absence of evidence for
large SOTs in bilayers with magnetic insulators we disregarded $G_{i}$ in our
previous work.\cite{Chiba14} However, there is no evidence against a strong
spin-orbit interaction at YIG$|$Pt interfaces either. The SMR phenomenology,
for example, can be also explained by an interface Rashba
interaction.\cite{Grigoryan14}

\begin{figure}[ptb]
\centering\includegraphics[width=0.43\textwidth,angle=0]{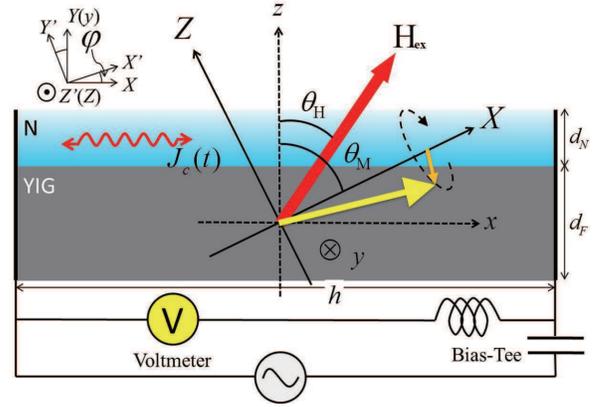}\caption{Schematic
set-up to observe the SMR rectified voltage in which $\mathbf{H_{\mathrm{ex}}%
}$ is an external magnetic field and $\theta_{\mathrm{H}}$ and $\theta
_{\mathrm{M}}$ show the external magnetic field and the magnetization angles.
The YIG($d_{F}$ nm)$|$N($d_{N}$ nm) bilayer film is patterned into a strip
with a length $h$. A Bias-Tee allows detection of a dc voltage under an ac
bias.}%
\label{oop}%
\end{figure}

In bilayer thin films made from a ferromagnetic and a normal metal, a dc
voltage is generated from the magnetization dynamics induced by an applied ac
spin-transfer torque and the anisotropic magnetoresistance (AMR). This
current-induced spin torque ferromagnetic resonance (ST-FMR) is an established
noninvasive method to study the spin-orbit coupling between currents and
magnetization.\cite{Liu} The spin-orbit coupling between currents and
magnetization in the YIG$|$Pt system can also be accessed in this manner by
utilizing the SMR as shown in Fig.~\ref{oop}.\cite{Chiba14}

Here we generalize our previous work on ST-FMR for bilayers of a ferro- or
ferrimagnetic insulator (FI) and a heavy normal metal (N)\cite{Chiba14} by
deriving magnetization dynamics and dc voltages for arbitrary equilibrium
magnetization directions and include heuristically a possibly large field-like
SOT in terms of a significant imaginary part of the spin-mixing conductance.

\section{Magnetization dynamics with spin-orbit torques}

The ac current with frequency $\omega_{a}=2\pi f_{a}$ induces a spin
accumulation distribution $\boldsymbol{\mu}_{s}(z,t)$ in N that fills the
spin-diffusion equation
\begin{equation}
\partial_{t}\boldsymbol{\mu}_{s}=D\partial_{z}^{2}\boldsymbol{\mu}_{s}%
-\frac{\boldsymbol{\mu}_{s}}{\tau_{\mathrm{sf}}}, \label{tdsd}%
\end{equation}
where $D$ is the charge diffusion constant and $\tau_{\mathrm{sf}}$ spin-flip
relaxation time in N with the spin-diffusion length $\lambda=\sqrt
{D\tau_{\mathrm{sf}}}$. The ISHE induces a charge current in the $x$-$y$ plane
by the diffusion spin current along the $z$-direction, $\overline{J_{c,x}}%
(t)$=$J_{\mathrm{SMR}}(t)+J_{\mathrm{SP}}(t)$ which is averaged over the N
film thickness $d_{N}$, where $J_{\mathrm{SMR}}(t)$ and $J_{\mathrm{SP}}(t)$
are SMR rectification and spin pumping-induced charge currents with $J_{c}%
^{0}(t)=J_{c}^{0}\operatorname{Re}(e^{i\omega_{a}t})$.\cite{Chiba14}

ST-FMR experiments utilize the ac impedance of the oscillating transverse spin
Hall current caused by the induced magnetization dynamics that is described by
the Landau-Lifshitz-Gilbert (LLG) equation including the interface spin
current $\boldsymbol{J}_{s}^{F|N}=\boldsymbol{J}_{s}^{T}+\boldsymbol{J}%
_{s}^{P}$ with
\begin{align}
\boldsymbol{J}_{s}^{T}  &  =\frac{G_{r}}{e}\hat{\mathbf{M}}\times\left(
\hat{\mathbf{M}}\times\boldsymbol{\mu}_{s}^{F|N}\right)  +\frac{G_{i}}{e}%
\hat{\mathbf{M}}\times\boldsymbol{\mu}_{s}^{F|N},\label{jst}\\
\boldsymbol{J}_{s}^{P}  &  =\frac{\hbar}{e}\left(  G_{r}\hat{\mathbf{M}}%
\times\partial_{t}\hat{\mathbf{M}}+G_{i}\partial_{t}\hat{\mathbf{M}}\right)  ,
\label{jsp}%
\end{align}
as the additional torques $\boldsymbol{\tau}_{J}=\gamma\hbar\boldsymbol{J}%
_{s}^{F|N}/(2eM_{s}d_{F})$ $(e=-|e|)$, where $\boldsymbol{\mu}_{s}^{F|N}$, $\hat
{\mathbf{M}}$, $\gamma$, $M_{s}$, and $d_{F}$ are the interface spin
accumulation between FI and N layers, the unit vector along the FI
magnetization, the gyromagnetic ratio, the saturation magnetization, and the
thickness of the FI film, respectively. The external magnetic field
$\mathbf{H_{\mathrm{ex}}}$ is applied at a polar angle $\theta_{\mathrm{H}}$,
especially in the $z$-$x$ plane corresponding to the angle $\alpha$ in Ref.~\onlinecite{Bai13}, 
and azimuth $\varphi$ in the $x$-$y$ plane. It is convenient to
consider the magnetization dynamics in the $(\theta_{\mathrm{M}}-\pi/2)$
around the $y$-axis and $\varphi$ around the $Z$-axis rotated coordinate
system [see Fig.~\ref{oop}]. Denoting the transformation matrix as
$R(\theta_{\mathrm{M}}-\pi/2,\varphi)$, the magnetization dynamics
$\mathbf{M}_{R}(t)=R(\theta_{\mathrm{M}}-\pi/2,\varphi)\mathbf{M}(t)$
precessing around the $Z^{\prime}$-axis obeys the LLG equation in the
$X^{\prime}Y^{\prime}Z^{\prime}$-coordinate system (Fig.~\ref{oop}),
\begin{equation}
\partial_{t}\hat{\mathbf{M}}_{R}=-\tilde{\gamma}\hat{\mathbf{M}}_{R}%
\times\left(  \mathbf{H}_{\mathrm{eff},R}+\mathbf{H}_{J,R}\right)  +\alpha
\hat{\mathbf{M}}_{R}\times\partial_{t}\hat{\mathbf{M}}_{R}, \label{llg}%
\end{equation}
where
\begin{align}
\mathbf{H}_{\mathrm{eff},R}  &  =\mathbf{H}_{\mathrm{ex}}+\mathbf{H}%
_{\mathrm{M}}+\mathbf{H}_{\mathrm{m}}(t)+\mathbf{H}_{\mathrm{ac}%
}(t)\nonumber\\
&  =%
\begin{pmatrix}
H_{\mathrm{ex}}\cos(\theta_{\mathrm{M}}-\theta_{\mathrm{H}})\\
0\\
H_{\mathrm{ex}}\sin(\theta_{\mathrm{M}}-\theta_{\mathrm{H}})
\end{pmatrix}
-4\pi M_{s}\cos\theta_{\mathrm{M}}%
\begin{pmatrix}
\cos\theta_{\mathrm{M}}\\
0\\
\sin\theta_{\mathrm{M}}%
\end{pmatrix}
\nonumber\\
&  -4\pi m_{Z}(t)\sin\theta_{\mathrm{M}}%
\begin{pmatrix}
\cos\theta_{\mathrm{M}}\\
0\\
\sin\theta_{\mathrm{M}}%
\end{pmatrix}
\nonumber\\
&  +%
\begin{pmatrix}
0\\
H_{ac}\cos\varphi\\
-H_{ac}\cos\theta_{\mathrm{M}}\sin\varphi
\end{pmatrix}
e^{i(\omega_{a}t+\delta)}%
\end{align}
is, respectively, the sum of the external magnetic field, the static
demagnetizing field, the dynamic demagnetization field, and the ac
current-induced Oersted field. $\delta$ is the phase shift between Oersted
field and current, which is governed by the details of the sample design and therefore
treated as an adjustable parameter.\cite{Harder11} The current-induced effective
field may be linearized
\begin{align}
\mathbf{H}_{J,R}  &  =(\hat{\mathbf{M}}_{R}\times H_{r}\hat{\mathbf{Y^{\prime}}%
}+H_{i}\hat{\mathbf{Y^{\prime}}})e^{i\omega_{a}t}\nonumber\\
&  \approx%
\begin{pmatrix}
0\\
H_{i}\cos\varphi+H_{r}\cos\theta_{\mathrm{M}}\sin\varphi\\
H_{r}\cos\varphi-H_{i}\cos\theta_{\mathrm{M}}\sin\varphi
\end{pmatrix}
e^{i\omega_{a}t},\\
&  H_{r(i)}=\frac{\hbar}{2|e|M_{s}d_{F}}\theta_{\mathrm{SH}}J_{c}^{0}%
\operatorname{Re}\left(  \operatorname{Im}\right)  \eta.
\end{align}
Here $\eta$ is the complex spin diffusion efficiency
\begin{equation}
\eta=\left(  1-\frac{1}{\cosh(d_{N}/\lambda)}\right)  \frac{\tilde{g}%
_{r}(1+\tilde{g}_{r})+\tilde{g}_{i}^{2}+i\tilde{g}_{i}}{(1+\tilde{g}_{r}%
)^{2}+\tilde{g}_{i}^{2}}%
\end{equation}
with $\tilde{g}_{r(i)}=2\lambda\rho G_{r(i)}\coth(d_{N}/\lambda)$ and $\rho$, the resistivity of bulk N . It is
plotted in Fig.~\ref{fig.VSMR} as a function of $r=d_{N}/\lambda$ and
$g_{r(i)}=2\lambda\rho G_{r(i)}$.
\begin{equation}
\alpha=\frac{\alpha_{0}+\beta\coth(r/2)\operatorname{Re}\eta}{1-\beta
\coth(r/2)\operatorname{Im}\eta}%
\end{equation}
is the modulated magnetization damping in terms of the Gilbert damping
constant of the isolated film $\alpha_{0}$ and $\beta=\gamma\hbar
^{2}/(4\lambda\rho e^{2}M_{s}d_{F})$, and $\tilde{\gamma}=\gamma/\left(
1-\beta\coth(r/2)\operatorname{Im}\eta\right)  $. The external magnetic field
$\mathbf{H}_{\mathrm{ex}}$ is applied at a polar angle $\theta_{\mathrm{H}}$
in the $z$-$x$ plane as shown in Fig.~\ref{oop}. It is convenient to consider
the magnetization dynamics in the $\theta_{\mathrm{M}}$ rotated coordinate
system in which the magnetization is stabilized along the static equilibrium
condition, $\mathbf{M}_{R}\times\mathbf{H}_{\mathrm{eff},R}=0$, obeys the relation
between $\theta_{\mathrm{H}}$ and $\theta_{\mathrm{M}}$ as
\begin{equation}
H_{\mathrm{ex}}=2\pi M_{s}\sin2\theta_{\mathrm{M}}/\sin(\theta_{\mathrm{M}%
}-\theta_{\mathrm{H}}).
\end{equation}

The magnetization precesses around the static equilibrium condition
$\mathbf{M}_{R}(t)=\mathbf{M}_{R}^{0}+\mathbf{m}_{R}(t)\approx\left(
M_{s},m_{Y^{\prime}}(t),m_{Z^{\prime}}(t)\right)  $, where $\mathbf{M}_{R}%
^{0}$ and $\mathbf{m}_{R}(t)$ are the static and dynamic components of the
magnetization. For a small-angle precession around the equilibrium direction
$\mathbf{M}_{R}^{0}$, $\mathbf{m}_{R}(t)=(0,\delta m_{Y^{\prime}}%
e^{i\omega_{a}t},\delta m_{Z^{\prime}}e^{i\omega_{a}t})$\ $\left(
\operatorname{Re}[\delta m_{Y^{\prime}}]\,\operatorname{Re}[\delta
m_{Z^{\prime}}]\ll M_{s}\right)  $, we linearize Eq.~(\ref{llg}) and arrive at
the FMR condition for the ac current frequency\cite{Ando11}
\begin{gather}
H_{\mathrm{FMR}}\cos(\theta_{\mathrm{M}}-\theta_{\mathrm{H}})=2\pi M_{s}\left(
\cos2\theta_{\mathrm{M}}+\cos^{2}\theta_{\mathrm{M}}\right) \nonumber\\
+\sqrt{\left\{  2\pi M_{s}\left(  \cos2\theta_{\mathrm{M}}-\cos^{2}%
\theta_{\mathrm{M}}\right)  \right\}  ^{2}+(\omega_{a}/\tilde{\gamma})^{2}}.
\end{gather}
The magnetization dynamics is
\begin{align}%
\begin{pmatrix}
m_{Y^{\prime}}(t)\\
m_{Z^{\prime}}(t)
\end{pmatrix}
&  =\frac{1}{2\pi}\frac{e^{i\omega_{a}t}\Omega_{a}\cos\varphi}{\Omega
_{\mathrm{H}}^{2}-\Omega_{a}^{2}+2i\tilde{\Delta}\Omega_{\mathrm{H}}}%
\begin{pmatrix}
Y_{r}^{c}+iY_{i}^{c}\\
Z_{r}^{c}+iZ_{i}^{c}%
\end{pmatrix}
\nonumber\\
&  +\frac{1}{2\pi}\frac{e^{i\omega_{a}t}\Omega_{a}\cos\theta_{\mathrm{M}}%
\sin\varphi}{\Omega_{\mathrm{H}}^{2}-\Omega_{a}^{2}+2i\tilde{\Delta}%
\Omega_{\mathrm{H}}}%
\begin{pmatrix}
Y_{r}^{s}+iY_{i}^{s}\\
Z_{r}^{s}+iZ_{i}^{s}%
\end{pmatrix}
,
\end{align}
where $\Omega_{\mathrm{H}}=\tilde{H}_{\mathrm{ex}}\cos(\theta_{\mathrm{M}}-\theta
_{\mathrm{H}})-\cos2\theta_{\mathrm{M}}-\cos^{2}\theta_{\mathrm{M}}$,
$\Omega_{a}=[\left(  \cos2\theta_{\mathrm{M}}-\cos^{2}\theta_{\mathrm{M}%
}\right)  ^{2}+\tilde{\omega}_{a}^{2}]^{1/2}$, $\tilde{\Delta}=\alpha
\tilde{\omega}_{a}$ is the linewidth, $Y_{r}^{c}=C_{+}\left(
H_{\mathrm{ac}}\cos\delta+H_{i}\right)  -C\alpha H_{\mathrm{ac}}\sin\delta$,
$Y_{i}^{c}=C\left[  H_{r}+\alpha\left(  H_{\mathrm{ac}}\cos\delta
+H_{i}\right)  \right]  +C_{+}H_{\mathrm{ac}}\sin\delta$, $Z_{r}^{c}%
=C_{-}H_{r}+CH_{\mathrm{ac}}\sin\delta$, $Z_{i}^{c}=C\left[  \alpha
H_{r}-\left(  H_{\mathrm{ac}}\cos\delta+H_{i}\right)  \right]  $, $Y_{r}%
^{s}=C_{+}H_{r}+CH_{\mathrm{ac}}\sin\delta$, $Y_{i}^{s}=C\left[  \alpha
H_{r}-\left(  H_{\mathrm{ac}}\cos\delta+H_{i}\right)  \right]  $, $Z_{r}%
^{s}=-C_{-}\left(  H_{\mathrm{ac}}\cos\delta+H_{i}\right)  +C\alpha
H_{\mathrm{ac}}\sin\delta$, $Z_{i}^{s}=-C\left[  H_{r}+\alpha\left(
H_{\mathrm{ac}}\cos\delta+H_{i}\right)  \right]  -C_{-}H_{\mathrm{ac}}%
\sin\delta$, $\tilde{\omega}_{a}=\omega_{a}/(2\pi M_{s}\tilde{\gamma})$,
$\tilde{H}_{\mathrm{ex}}=H_{\mathrm{ex}}/(2\pi M_{s})$,
$C=\tilde{\omega}_{a}/\Omega_{a}$, and $C_{+(-)}=1+(-)(\cos^2\theta_{\mathrm{M}}-\cos2\theta_{\mathrm{M}})/\Omega_{a}$.

\section{SMR rectification and spin pumping voltages}

\begin{figure}[ptb]
\includegraphics[width=0.45\textwidth,angle=0]{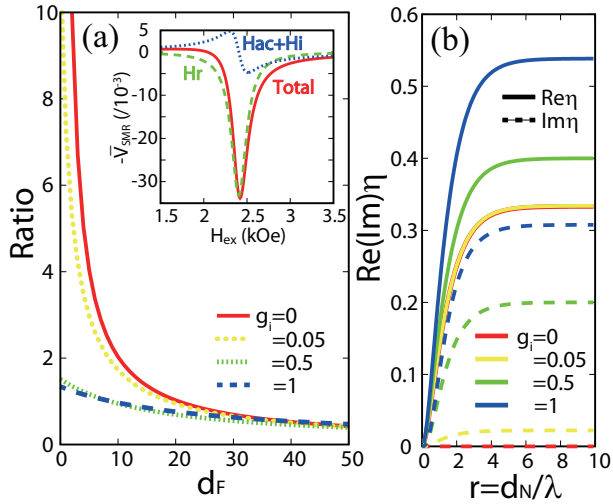}\caption{(a)
Spin diffusion efficiency $\operatorname{Re}\left(  \operatorname{Im}\right)
\eta$ as a function of $r=d_{N}/\lambda$ for $g_{r}=0.5$ and different $g_{i}$. 
(b) The YIG thickness dependence of the ratio of symmetric and antisymmetric
contributions to the rectified voltage in YIG$|$N with $g_{r}=0.5$,
$\theta_{\mathrm{SH}}=0.1$, $r=4$, $\lambda=1.5$ nm, and $M_{s}=1.56\times
10^{5}\ \mathrm{A/m}$ at $f_{a}=9\,$GHz. Insets represent the calculated SMR
rectified voltage $\bar{V}_{\mathrm{SMR}}$ normalized by $(h\Delta\rho
_{1}J_{c}^{0}/4)\cos\varphi\sin2\varphi$ for $g_{i}=0.05$ and $\beta/\alpha_0=1500$.}%
\label{fig.VSMR}%
\end{figure}

In the ST-FMR measurement, the dc voltage arises from the mixing of the
applied ac current and the oscillating SMR in N (spin rectification) as well
as the ISHE mediated spin pumping. This method is analogous to electrical
detection of FMR in which the magnetization dynamics is excited by microwaves
in coplanar wave guides or cavities. Here we focus on the current-induced
magnetization dynamics which induces down-converted dc (and second harmonic)
components in the N layer. Denoting the time average by $\left\langle
\cdots\right\rangle _{t}$, the open-circuit dc voltage is $V_{DC}=h\rho
\langle\overline{J_{c,x}}(t)\rangle_{t}=V_{\mathrm{SMR}}+V_{\mathrm{SP}}$. The dc voltages due to SMR
rectification and spin pumping are \begin{widetext}
\begin{align}
V_{\mathrm{SMR}}
&= \frac{h\varDelta\rho_{1}J_{c}^{0}}{4}\frac{F_{S}(\tilde{H}_{\mathrm{ex}}
)}{\tilde{\Delta}}\left(  \tilde{Y}_{i}^c+ \tilde{Y}_{r}^c\frac{\tilde{H}_{\mathrm{ex}}-\tilde{H}_{\mathrm{FMR}}}{\tilde{\Delta}}\cos(\theta_{\mathrm{M}}-\theta_{\mathrm{H}})\right)
\cos\varphi\sin2\varphi\sin\theta_{\mathrm{M}}\nonumber\\
&\ \ \ -\frac{h\varDelta\rho_{1}J_{c}^{0}}{4}\frac{F_{S}(\tilde{H}_{\mathrm{ex}}
)}{\tilde{\Delta}}\left(  \tilde{Z}_{i}^s+ \tilde{Z}_{r}^s\frac{\tilde{H}_{\mathrm{ex}}-\tilde{H}_{\mathrm{FMR}}}{\tilde{\Delta}}\cos(\theta_{\mathrm{M}}-\theta_{\mathrm{H}})\right)
\sin^3\varphi\cos\theta_{\mathrm{M}}\sin2\theta_{\mathrm{M}}\nonumber\\
&\ \ \ +\frac{h\varDelta\rho_{1}J_{c}^{0}}{8}\frac{F_{S}(\tilde{H}_{\mathrm{ex}}
)}{\tilde{\Delta}} \left(\tilde{Y}_{r}^s-\tilde{Z}_{r}^c\right)\frac{\tilde{H}_{\mathrm{ex}}-\tilde{H}_{\mathrm{FMR}}}{\tilde{\Delta}}\cos(\theta_{\mathrm{M}}-\theta_{\mathrm{H}})
\sin\varphi\sin2\varphi\sin2\theta_{\mathrm{M}}, \label{VSMR}\\
\nonumber\\
V_{\mathrm{SP}}
&= \frac{h\rho J_{r}^{P}}{4}\frac{F_{S}(\tilde{H}_{\mathrm{ex}}
)}{\tilde{\Delta}^2}\left(  \tilde{Z}_{i}^c\tilde{Y}_{r}^c- \tilde{Z}_{r}^c\tilde{Y}_{i}^c\right)\cos\varphi\sin2\varphi\sin\theta_{\mathrm{M}}\nonumber\\
&\ \ \ +\frac{h\rho J_{r}^{P}}{4}\frac{F_{S}(\tilde{H}_{\mathrm{ex}}
)}{\tilde{\Delta}^2}\left(  \tilde{Z}_{i}^s\tilde{Y}_{r}^s- \tilde{Z}_{r}^s\tilde{Y}_{i}^s\right)\sin^3\varphi\cos\theta_{\mathrm{M}}\sin2\theta_{\mathrm{M}}\nonumber\\
&\ \ \ +\frac{h\rho J_{r}^{P}}{8}\frac{F_{S}(\tilde{H}_{\mathrm{ex}}
)}{\tilde{\Delta}^2}\left(  \tilde{Z}_{i}^c\tilde{Y}_{r}^s- \tilde{Z}_{r}^c\tilde{Y}_{i}^s+\tilde{Z}_{i}^s\tilde{Y}_{r}^c- \tilde{Z}_{r}^s\tilde{Y}_{i}^c\right)\sin\varphi\sin2\varphi\sin2\theta_{\mathrm{M}}, \label{VSP}
\end{align}
\end{widetext}where $\Delta\rho_{1}\propto\theta_{\mathrm{SH}}^{2}%
\operatorname{Re}\eta$ is the conventional dc-SMR,\cite{Chen,Chiba14}
$J_{r}^{P}$=$\hbar\omega_{a}/(2|e|d_{N}\rho)\theta_{\mathrm{SH}}%
\operatorname{Re}\eta$, $F_{S}(\tilde{H}_{\mathrm{ex}})=\tilde{\Delta}^{2}/[(
\tilde{H}_{\mathrm{ex}}-\tilde{H}_{\mathrm{FMR}})  ^{2}\cos^{2}(\theta_{\mathrm{M}}-\theta
_{\mathrm{H}})+\tilde{\Delta}^{2}]$, $\tilde{Y}_{r(i)}^{c.s}=Y_{r(i)}^{c.s}/(2\pi
M_{s})$, and $\tilde{Z}_{r(i)}^{c.s}=Z_{r(i)}^{c.s}/(2\pi M_{s})$. 
The third term of Eq.~(\ref{VSMR}) is directly proportional to 
$\tilde{Y}_{r}^s-\tilde{Z}_{r}^c=(C_{+}-C_{-})H_{r}$, thereby being independent of the as yet unknown $\delta$,
which can be helpful in picking up the purely spin-torque induced FMR. 
For the angle $\varphi=0$ (external magnetic field in $zx$-plane),\cite{Bai13} the
SMR-rectified voltage vanishes since the SMR oscillates with twice the
frequency $e^{2i\omega_{a}t}$ (Fig.~\ref{oop}), illustrating that the SMR is
essentially different from the AMR in feromagnetic metals. To study SOTs by
ST-FMR, the in-plane ($\theta_{\mathrm{M}}=\pi/2$) configuration is therefore
sufficient (provided $\delta=0$). The ratio between symmetric and
antisymmetric components in Eq.~(\ref{VSMR}) is
\[
\mathrm{Ratio}=\frac{C}{C_{+}}\left[  \operatorname{Re}\eta\left(  \frac
{4\pi^{2}M_{s}d_{F}d_{N}}{\theta_{\mathrm{SH}}\Phi_{0}}+\operatorname{Im}%
\eta\right)  ^{-1}+\alpha\right]  ,
\]
where $\Phi_{0}$ is the flux quantum. The calculated ratio is plotted in
Fig.~\ref{fig.VSMR} as a function of the FI layer thickness $d_{F}$ while
the inset shows $V_{\mathrm{SMR}}$ including the symmetric
contributions by the spin transfer torque as well as the antisymmetric ones by
the Oersted magnetic field and the field-like SOT. This ratio (and dc voltages
itself) depends sensitively on $d_{F}$ since the Gilbert damping in YIG is
very weak.

\section{SUMMARY AND DISCUSSIONS}

In summary, we present a theory of the ac current-driven ST-FMR in bilayer
systems made from a magnetic insulator such as YIG and a heavy metal such as
Pt with emphasis on the following two points: (i) expressions for the dc
voltage for all  directions and strengths of the applied magnetic field and
(ii) the magnetization dynamics in the presence of a field-like spin-orbit
torque. The dc voltages generated in YIG$|$N bilayers are found to depend
sensitively on the ferromagnet layer thickness when the bulk Gilbert damping
is small. For thin YIG layers the line shape can be significantly affected by
the imaginary part of spin-mixing conductance through the field-like
spin-orbit torque. Thermal effects such as the spin Seebeck effect caused by
Joule heating in N may contribute to the SMR rectified voltage only in the
form of a constant background dc voltage. Our predictions can be tested
experimentally by ST-FMR experiments with a magnetic insulator that would
yield valuable insights into the conduction electron spin-interface exchange
interaction and spin-orbit coupling between currents and magnetization at the
interface of magnetic insulators and metals.


We would like to thank Can-Ming Hu for stimulating
communications. This work was supported by KAKENHI (Grants-in-Aid for
Scientific Research) Nos. 22540346, 25247056, 25220910, and 268063, FOM
(Stichting voor Fundamenteel Onderzoek der Materie),\ the ICC-IMR, EU-FET grant InSpin 612759, and DFG Priority Programme 1538
\textquotedblleft{Spin-Caloric Transport}\textquotedblright\ (Grant No. BA 2954/1).

\end{document}